\documentstyle[aps,epsfig,multicol]{revtex}

\begin{document}

\title{Absence of gross inhomogeneity in HTS cuprates}

\author{J.W. Loram$^1$, J.L. Tallon$^2$ and W.Y. Liang$^1$}

\address{$^1$Interdisciplinary Research Center in Superconductivity,
Cambridge University, Cambridge CB3 0HE, England.}
\address{$^2$MacDiarmid Institute for Advanced Materials and
Nanotechnology, Industrial Research Ltd., and Victoria University,
P.O. Box 31310, Lower Hutt, New Zealand.}

\date{\today}

\maketitle

\begin{abstract}
Recent scanning tunnelling microscopy (STM) studies on
Bi$_2$Sr$_2$CaCu$_2$O$_{8+\delta}$ (Bi-2212) revealed the presence
of severe inhomogeneity with length scale $L_0 \approx \xi_0$, the
coherence length. Other studies have been interpreted in terms of
mesoscale or nanoscale phase segregation. Here we analyze heat
capacity and NMR data on Bi-2212 and
(Y,Ca)Ba$_2$Cu$_3$O$_{7-\delta}$ and find no evidence for phase
segregation or gross inhomogeneity. For Bi-2212 the coherence
scale $L_0$ increases with doping from 5 to 17$\xi_0$ and the hole
density inhomogeneity decreases from 0.028 to 0.005 holes/Cu. We
conclude that STM measurements considerably overstate the
inhomogeneity in bulk Bi-2212.
\end{abstract}


\begin{multicols}{2}

The possible importance of inhomogeneity in the physics of high
temperature superconducting (HTS) cuprates has been stressed since
the earliest investigations of these materials. Of necessity they
involve chemical disorder as substitutional dopants in the parent
compounds are usually required to introduce carriers. Frequently
observed forms of inhomogeneity include substitutional,
interstitial, reconstructive, twinning and
incommensuration\cite{Stoto,Tallon1,Tarascon,Subra}. Despite this
potentially high degree of structural and compositional disorder
(which is quite variable from one compound to another) the
cuprates display universal phase
behaviour\cite{Tallon2,Loram1,Talloram}. Their physical properties
vary systematically with doping in very much the same way from one
compound to another provided simply that the substitutional
disorder lies outside of the active CuO$_2$ planes. This, of
course, reflects the strongly two-dimensional nature of the
electron dynamics in the CuO$_2$ planes. More fundamentally
however, inhomogeneity of intrinsic electronic origin has been
discussed widely, most notably various nanoscale stripe phases in
which the background spins separate spatially from doped charges
to form quasi-one-dimensional structures\cite{Stripes1}. There is
thus much scope in HTS materials for significant extrinsic and
intrinsic spatial inhomogeneity.

Recent high-resolution scanning tunnelling microscope (STM)
studies\cite{Pan,Lang} at 4.2K have provided dramatic evidence of
electronic inhomogeneity on the surfaces of Bi-2212 in the
superconducting (SC) state. These reveal a patchwork of
contrasting regions with length scale $\approx$ 14$\AA$,
comparable with the coherence length $\xi_0$, some having
well-defined coherence peaks and others having strongly-reduced
coherence peaks shifted to higher energies similar to the features
of the pseudogap above $T_c$\cite{Renner}. The $\sim$ 50\% spread
in the magnitude of the SC energy gap, $\Delta_0$, was shown to
correlate with a similar spread in the local density of states.
These variations were attributed to a distribution of hole
concentration, $p$, arising from oxygen disorder and a FWHM spread
of $\triangle p \approx$ 0.08 holes/Cu was inferred by comparing
the distribution in SC gap with the $p$-dependence of  $\Delta_0$
deduced from ARPES. It was suggested that such inhomogeneity may
be quite general and may reflect "an intimate relationship with
superconductivity" in the cuprates\cite{Pan}. However, it cannot
be presumed that such inhomogeneity is generic nor that it even
reflects the nature of the $bulk$ electronic state. STM is
strictly a surface probe susceptible to just the outermost CuO$_2$
layer. Little is known about surface reconstruction in this
material, the location of additional oxygen atoms in the cleaved
BiO layer or the role of Bi atoms incorporated into the underlying
SrO layer. We argue here from thermodynamic measurements and NMR
data that the bulk electronic state is much more homogeneous than
is suggested by the STM results which, by implication, are an
artifact of a perturbed surface.

A quantitative measure of the degree of doping inhomogeneity in
the HTS cuprates can be inferred from NMR linewidths. NMR is a
bulk probe which is sensitive to the local electronic state near
the probe nucleus. We show in Fig. 1 the $^{89}$Y Knight shift,
$^{89}K_s$, at 300K for
Y$_{0.8}$Ca$_{0.2}$Ba$_2$Cu$_3$O$_{7-\delta}$ (Y,Ca-123) at six
different values of $\delta$ and $p$ spanning the SC phase
diagram\cite{Williams1}. The $p$ values were obtained from
thermopower measurements and $T_c$ values are also plotted. As is
well known, the Knight shift varies systematically with doping and
the figure reveals a slope of 580 ppm/hole. Doping inhomogeneity
will therefore induce a $minimum$ line width of 580 $\times
\triangle p$ ppm. Other contributions, such as local
moments\cite{Alloul} and the effect of disorder on the chemical
shift will also add to the linewidth. We may thus use the observed
linewidth to determine the $maximum$ doping inhomogeneity (if all
the linewidth came only from $\triangle p$).

The vertical bands in the figure show the FWHM in the hole
concentration determined in this way\cite{Tallon3}. We have used
the magic-angle-spinning data of Balakrishnan {\em et
al.}\cite{Bala} for YBa$_2$Cu$_3$O$_{7-\delta}$ (Y-123) and of
Williams {\em et al.}\cite{Williams2} for YBa$_2$Cu$_4$O$_8$
(Y-124). In the last case the linewidth is extremely narrow, about
100 Hz, giving $\triangle p\leq$ 0.0069. It is clear that this
class of HTS is remarkably homogeneous yet they exhibit the same
generic thermodynamic and ground-state behavior as Bi-2212\cite{Loram2}.

\begin{figure}
\centerline{\includegraphics*[height=50mm,
width=70mm]{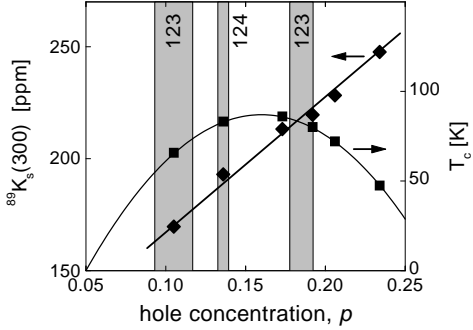}} \caption{\small The doping dependence of
the $^{89}$Y Knight shift at 300K for
Y$_{0.8}$Ca$_{0.2}$Ba$_2$Cu$_3$O$_{7-\delta}$. From this the NMR
line-width is used to estimate a maximum FWHM, $\Delta$$p$, in
doping state shown for Y-123 and Y-124 by the vertical bands.}
\end{figure}

Heat capacity studies on HTS cuprates provide a number of
potential checks against electronic inhomogeneity. These include
(i) the value of the specific heat coefficient, $\gamma \equiv
C_p/T$ at $T$=0, (ii) its value, $\gamma_n$, at high temperature
in the normal state, (iii) the magnitude and doping dependence of
the jump, $\triangle\gamma$, in $\gamma$ at $T_c$ and (iv) the
transition width. For example, mesoscale segregation into SC and
normal metallic phases will result in normal states in the gap
manifested by a non-zero value of $\gamma$ at $T$=0. Progressive
segregation with overdoping, for example, will see $\gamma$(0)
rising steadily, filling out the $d$-wave gap. On the other hand,
mesoscale segregation into insulating and normal metallic phases
will result in a progressive reduction in the high-temperature
value of $\gamma_n$ as states are removed. Here we focus on the
thermodynamic data for Bi-2212 because this is the system in which
the STM inhomogeneity was observed. To this we add data for
Y,Ca-123, a very different system, to show that our observations
are generic to the cuprates.

Ceramic samples {($\approx$ 2 gm)} of
Bi$_{2.1}$Sr$_{1.9}$CaCu$_2$O$_{8+\delta}$\cite{Loram6} and
Y$_{0.8}$Ca$_{0.2}$Ba$_2$Cu$_3$O$_{7-\delta}$\cite{Loram7} were
synthesized using standard solid-state reaction and were oxygen
loaded by slow cooling and annealing for many days at
325$^{\circ}$C in flowing oxygen. The electronic heat capacity was
measured using a high-precision differential technique described
elsewhere\cite{Loram3}. The samples were then alternately annealed
and quenched into liquid nitrogen and the heat capacity remeasured
so as to track the thermodynamic state as a function of
progressively depleted oxygen content. Changes in oxygen
stoichiometry were determined from the mass changes and these
concurred rather precisely with changes in the phonon specific
heat. Values of $p$ were determined from $T_c$ values using the
approximate parabolic phase curve, $T_c(p)$, given
by\cite{Tallon2}
\begin{equation}
\ T_c(p) = T_{c,max}~{\bf [}~1-82.6~(p-0.16)^2~{\bf ]}.
\end{equation}
Similar studies have correlated such values of $p$ with the
thermopower for both Bi-2212\cite{Ober} and
(Y,Ca)-123\cite{Bernhard}.

Fig. 2 shows a series of $\gamma(T)$ curves for each sample at
closely spaced $p$-values from deeply underdoped to deeply
overdoped. Dashed curves denote optimal doping where $T_c$
maximises while the thick curves denote critical doping
($p_{crit}$=0.19) where the pseudogap abruptly closes. Apart from
small impurity-induced upturns for $T <$ 10 K $\gamma(0)$ remains
close to zero across the entire doping range
0.084$\leq$p$\leq$0.23 in Y-123 and 0.13$\leq$p$\leq$0.21 in
Bi-2212, while the values of $\gamma(T)$ at high temperature are
constant, independent of doping and characteristic of a uniform
normal metal. These observations rule out the two types of
mesoscale segregation discussed above. We note that the $^{17}$O
Knight shift in overdoped Y,Ca-123 also approaches zero at
$T$=0\cite{Williams1} in agreement with the thermodynamic data and
independently confirming the absence of unpaired quasiparticles.
Significant gap filling in the single layer cuprates
Tl-2201\cite{Loram4} and La-214\cite{Loram5} only occurs in the
most heavily overdoped region p$\geq$0.24. This may reflect the
increased role of pair breaking relative to a rapidly decreasing
superconducting gap, and does not necessarily indicate
segregation.

\begin{figure}
\centerline{\includegraphics*[height=65mm,
width=60mm]{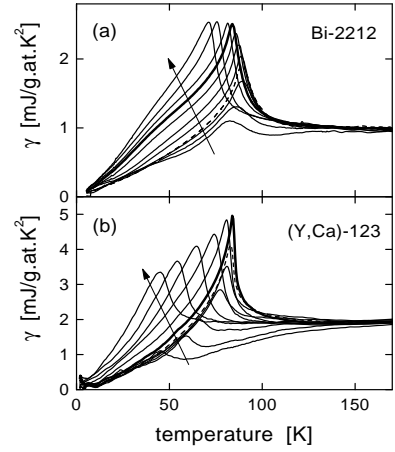}} \caption{\small The specific heat
coefficient, $\gamma(T)$, for a series of doping states for
Bi-2212 (0.13 $\leq p \leq$ 0.21) [17] and (Y,Ca)-123 (0.084 $\leq
p \leq$ 0.23) [18]. Arrows indicate increasing doping, bold curves
denote critical doping, $p$ = 0.19 and dashed curves optimal
doping, $p$ = 0.16.}
\end{figure}

Fig. 2 also reveals rather sudden changes in $\gamma(T,p)$ with
doping. The jump, $\triangle\gamma$, at the transition changes
only slowly across the deeply overdoped region but, starting at
$p_{crit}$=0.19, the magnitude of $\triangle\gamma$ falls very
rapidly with further reduction in doping. Adjacent curves are
separated by about $\triangle p\approx$0.008 and the observed
sudden collapse of the specific heat anomaly suggests that any
spread in doping state arising from nanoscale inhomogeneity must
be less than this value. Consistent with this inference
the transition widths remain quite narrow and we now focus on this
issue in more detail.

A finite transition width $\triangle t_m$ may arise from a spread,
$\triangle p$, in doping over $mesoscopic$ regions (due for
example to a spread, $\triangle x$, in local donor density, $x$).
Here,
\begin{equation}
\ \triangle t_m = \triangle T_c/T_c = T_c^{-1} \mid dT_c/dp\mid
\times \triangle p,
\end{equation}
which falls to a minimum at optimum doping $p$ = $p_{opt}$ = 0.16.
On the other hand a finite transition width $\triangle t_{fs}$
independent of $\mid dT_c/dp\mid$ will result from inhomogeneity
on a much shorter length scale, $\xi$ the coherence length, due to
$finite-size$ effects \cite{Fisher,Inderhees,Schneider}. In the
thermodynamic limit the coherence length, $\xi(T)$, diverges as $t
= T/T_c - 1 \rightarrow 0$ according to the power law $\xi(T)$ =
$\xi_0/\mid$$t\mid ^\nu$, where $\nu$=0.67 ($\approx$ 2/3) for
3D-XY fluctuations. Where the system is inhomogeneous this
divergence is cut off by a length scale $L_0$ which is presumed by
several authors\cite{Fisher,Inderhees,Schneider} to be the spatial
length scale of the inhomogeneity. As a consequence the divergence
of critical fluctuations is suppressed within
\begin{equation}
\ \triangle t_{fs} = (\xi_0/L_0)^{3/2}.
\end{equation}
It is immediately apparent that, if $L_0 \approx \xi_0$ as
inferred from STM\cite{Pan,Lang}, then the transition broadening
must be comparable to the magnitude of $T_c$, which is clearly not
the case. In fact, typical transition widths of $\approx$ 2-3K
require that $L_0 / \xi_0 >$ 10. For any doping state we can
estimate the transition width from the finite cut-off of the
fluctuation-induced heat capacity as $\mid t\mid \rightarrow 0$
and we will see that it is indeed small.

\begin{figure}
\centerline{\includegraphics*[height=80mm,
width=70mm]{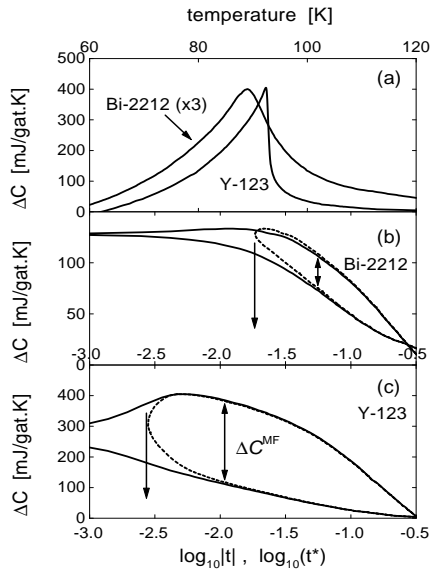}} \caption{\small (a) The specific heat
anomaly near $T_c$ for Bi-2212 and YBa$_2$Cu$_3$O$_{6.92}$, both
with $p\approx$ 0.18. Panels (b)and (c) show $\Delta C$ versus
$\log_{10}\mid$$t\mid$ (solid curve) and versus $\log_{10}$($t$*)
(dashed curve), where $t^* = (t^2 + \triangle t^2)^{1/2}$.
Single-head arrows show the transition width, $\Delta t$, and
double-head arrows show the mean-field jump in specific heat,
$\Delta C^{MF}$.}
\end{figure}

Fig. 3(a) shows the specific heat near $T_c$ of Bi-2212 and pure
YBa$_2$Cu$_3$O$_{6.92}$, both at a similar doping state $p\approx
0.18$. In the 3D-XY model the logarithmic divergence of the
specific heat near $T_c$ is given by $\Delta C_p$ = $A^-\ln\mid
t\mid$ + $\triangle C_p^{MF}$ for $t<$0 and $A^+ \ln\mid t\mid$
for $t>$0. Here $\triangle C_p^{MF}$ is the mean field step, $A^-
\approx A^+ = 4k_B/(9\pi^2\Omega(0))$ and $\Omega$ is the
coherence volume\cite{Kulic}. A semilog plot of $\triangle C_p$
versus $\ln\mid$$t\mid$ gives two parallel lines offset by the
mean-field step, $\triangle C_p^{MF}$. In practice this plot
(shown in Fig. 3(b) for Bi-2212 and in Fig. 3(c) for Y-123)
exhibits negative curvature for sufficiently small $\mid$$t\mid$
due to the finite-size cut-off in critical fluctuations or other
sources of broadening. The effect of a spread in $T_c$ may be
modelled by replacing t by $t^* = (t^2 + \triangle t^2)^{1/2}$ in
the above expressions for $\triangle C_p$ \cite{Inderhees}. This
is illustrated in Figs. 3(b) and (c) by selecting values of the
half-width $\triangle t$ that just avert negative curvature. Since
$\triangle t_{fs}$ and $\triangle t_m \leq \triangle t $ the
measured broadening $\triangle t$ yields a lower limit to
$L_0/\xi_0$ and an upper limit to $\triangle p$ (eqs. 2,3). From
the figures one finds $\triangle t$ = 0.018 and $\triangle
C_p^{MF}$ = 34 mJ/g.at.K for Bi-2212; and $\triangle t$ = 0.0028
and $\triangle C_p^{MF}$ = 266 mJ/g.at.K for
YBa$_2$Cu$_3$O$_{6.92}$. These values for $\triangle t$ translate
to $L_0 \geq$ 16 $\xi_0$ and 50 $\xi_0$, respectively, values
which are very much larger than those implied by the STM results.

\begin{figure}
\centerline{\includegraphics*[height=80mm,
width=80mm]{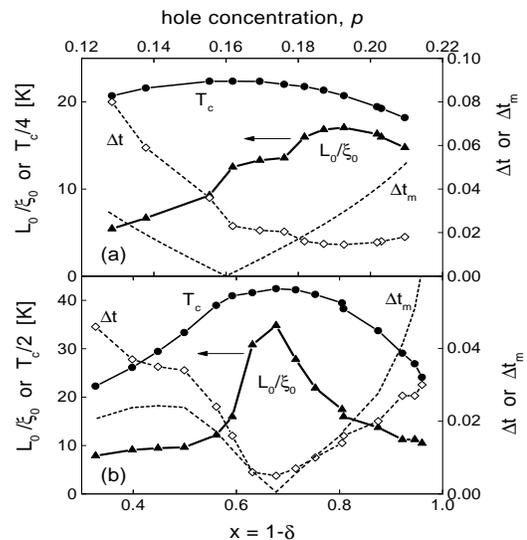}} \caption{\small The half-width of the
transition $\Delta t$ and lower limits for $L_0/\xi_0$ derived
from eq. (3) for (a) Bi-2212 and (b) (Y,Ca)-123. Values of $\Delta
t_m$ calculated using eq. (2) are shown in (a) for $\Delta p$ =
0.005 and in (b) for $\Delta x$ = 0.01.}
\end{figure}

The same analysis may be carried out for each investigated doping
level in Y$_{0.8}$Ca$_{0.2}$Ba$_2$Cu$_3$O$_{7-\delta}$ and in
Bi-2212. Fig. 4 shows  the measured broadening $\triangle t$ and
the corresponding lower limit to $L_0/\xi_0$ plotted as a function
of $p$ in the case of Bi-2212 and of oxygen content, $x=1-\delta$,
for (Y,Ca)-123. In the former case lower-limit values of
$L_0$/$\xi_0$ range from 17 at critical doping ($p_{crit}$ = 0.19)
down to 5 in the most heavily underdoped samples. For (Y,Ca)-123,
though less than for pure Y-123, values of $L_0$/$\xi_0$ are still
as high as 35. Even in the most extreme cases the length scale is
almost an order of magnitude longer than that inferred from the
STM studies. We also show in Fig. 4 $\Delta t_m$ obtained from eq.
(2) using $\Delta p$ = 0.005 for Bi-2212 and $\Delta x$ = 0.01 for
(Y,Ca)-123. In the latter case $\Delta t_m$ follows $\triangle t$
rather well across the phase diagram indicating a spread in oxygen
content of $\triangle x\approx \pm$0.01 for this compound. In
Bi-2212 the transition width $\triangle t$ does not correlate with
$\Delta t_m$, confirming the existence of short-length-scale
inhomogeneity. Nevertheless we can conclude that the spread in p
does not exceed $\approx$ 0.005 at p= 0.21 and $\approx$ 0.028 at
p=0.13, significantly less than the value inferred from the STM
data.

We plot in Fig. 5 the magnitude of the deduced mean-field step,
$\triangle C_p^{MF}$, for Bi-2212. This remains rather constant on
the overdoped side but falls extremely sharply at $p_{crit}$ =
0.19 as the pseudogap opens and, remarkably, is essentially zero
by optimal doping. A mere change of $\Delta p$ = 0.03 results in
the complete collapse to zero of the mean-field step. This abrupt
change again underscores the degree of bulk electronic homogeneity
that must be present, even in the case of Bi-2212. With $\triangle
C_p^{MF} \approx$ 0 for $p \leq$ 0.16 the specific heat anomaly
becomes symmetric about $T_c$ because its weight arises $purely$
from critical fluctuations. We note that there is no significance
in the fact that the mean-field step reaches zero at $p_{opt}$.
This is not the case in Y-123 and we expect that the collapse in
$\triangle C_p^{MF}$ would be even more abrupt in a system with
weaker interlayer coupling than Bi-2212. The coherence volume,
$\Omega (0)$, is also shown in the figure. The values are
consistent with $\xi_{ab}\approx$ 17 $\AA$ and $\xi_c\approx$ 0.5
$\AA$.

\begin{figure}
\centerline{\includegraphics*[height=50mm,
width=70mm]{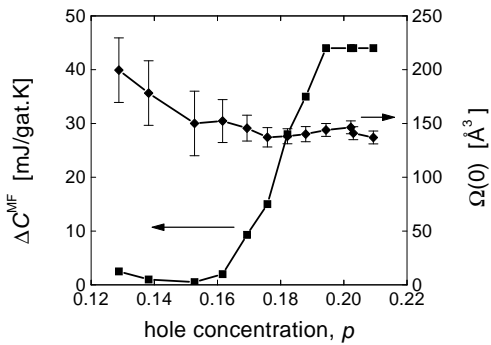}} \caption{\small The doping dependence of
the mean field jump, $\Delta C^{MF}$, in specific heat for
Bi$_{2.1}$Sr$_{1.9}$CaCu$_2$O$_{8+\delta}$ determined from the
fluctuation analysis as in Fig. 2(b) for each annealing state. The
coherence volume $\Omega (0)$ from the amplitude $A^+ =
4k_B/(9\pi^2\Omega(0))$ is also shown.}
\end{figure}

We conclude that specific heat and NMR measurements indicate
levels of inhomogeneity much lower than those inferred from STM
experiments. In the doping range considered there are no
phase-separated "normal" or "insulating" regions coexisting with
superconductivity. Y-123 and Bi-2212 reveal very narrow
thermodynamic features, such as the jump in $\gamma \equiv C_p/T$,
which change abruptly with doping indicating a spread in local
hole concentration no more than $\triangle p \approx \pm$0.01.
Such features, found also in
Y$_{1-x}$Ca$_x$Ba$_2$Cu$_3$O$_{7-\delta}$,
Tl$_{0.5}$Pb$_{0.5}$Sr$_2$Ca$_{1-x}$Y$_x$Cu$_2$O$_7$ and in
La$_{2-x}$Sr$_x$CuO$_4$, are generic to the cuprates and suggest
(at least for $p\geq$0.125) a remarkable degree of bulk electronic
homogeneity in spite of considerable disorder in the non-CuO$_2$
layers. Analysis of critical fluctuations near $T_c$ indicates a
spread of local donor densities no greater than $\triangle x
\approx \pm$0.01 for Y$_{1-x}$Ca$_x$Ba$_2$Cu$_3$O$_{7-\delta}$ and
$\approx \pm$0.02 for Bi-2212, while the latter compound exhibits
a sudden and complete collapse of the mean-field step between
critical and optimal doping. We conclude that the static
inhomogeneity observed in STM on Bi-2212 single crystals is
unrelated to bulk superconducting HTS properties and probably
reflects the presence of disorder at the surface.

% \begin{figure}
% \includegraphics{}%
% \caption{\label{}}
% \end{figure}

% Surround figure environment with turnpage environment for landscape
% figure
% \begin{turnpage}
% \begin{figure}
% \includegraphics{}%
% \caption{\label{}}
% \end{figure}
% \end{turnpage}

This work was supported by the Engineering and Physical Sciences
Research Council (JWL \& WYL) and the Marsden Fund (JLT).

\end{multicols}
\end{document}